\documentclass[%
preprint,
 amsmath,amssymb,
aps,
prb,
showkeys,
endfloats*,
multicol,
]{revtex4-1}

\usepackage{graphicx}
\usepackage{dcolumn}
\usepackage{bm}
\usepackage{color}
\usepackage{txfonts}

\newcommand{\vecvar}[1]{\mbox{\boldmath$#1$}}
\newcommand{\e}{\mbox{e}}

\begin{document}
\preprint{CCS9001}

\title[]{Real-space method for first-principles electron-transport calculations: self-energy terms of electrodes for large systems}

\author{Tomoya Ono}%
\email{ono@ccs.tsukuba.ac.jp}
\affiliation{Center for Computational Sciences, University of Tsukuba, Tsukuba, Ibaraki 305-8577, Japan}%
\affiliation{JST-PRESTO, Kawaguchi, Saitama 332-0012, Japan}%
\author{Shigeru Tsukamoto}
\affiliation{Peter Gr\"{u}nberg Institut \& Institute for Advanced Simulation, Forschungszentrum J\"{u}lich and JARA, D-52425 J\"{u}lich, Germany}

\date{\today}

\begin{abstract}
We present a fast and stable numerical technique to obtain the self-energy terms of electrodes for first-principles electron-transport calculations. Although first-principles calculations based on the real-space finite-difference method are advantageous for execution on massively parallel computers, large-scale transport calculations are hampered by the computational cost and numerical instability of the computation of the self-energy terms. Using the orthogonal complement vectors of the space spanned by the generalized Bloch waves that actually contribute to transport phenomena, the computational accuracy of transport properties is significantly improved with a moderate computational cost. To demonstrate the efficiency of the present technique, the electron-transport properties of a Stone-Wales (SW) defect in graphene and silicene are examined. The resonance scattering of the SW defect is observed in the conductance spectrum of silicene since the $\sigma^\ast$ state of silicene lies near the Fermi energy. In addition, we found that one conduction channel is sensitive to a defect near the Fermi energy, while the other channel is hardly affected. This characteristic behavior of the conduction channels is interpreted in terms of the bonding network between the bilattices of the honeycomb structure in the formation of the SW defect. The present technique enables us to distinguish the different behaviors of the two conduction channels in graphene and silicene owing to its excellent accuracy.
\end{abstract}

\pacs{71.15.-m, 72.10.-d, 72.80.Vp, 73.40.-c}
\maketitle
\section{Introduction}
\label{sec:introduction}
Recently, quantum-transport calculations have become an important tool for investigating the physics and chemistry of nanoscale systems because they are expected to exhibit considerably different transport properties from those of classical conductors. Owing to the complexity of the problem, such studies are strongly dependent on the existence of reliable numerical treatments based on first-principles approaches. A number of first-principles methods for calculating the electron-transport properties of nanoscale systems have been proposed so far. They are roughly categorized into two approaches. One approach uses the nonequilibrium Green's function (NEGF). The relation between the conductance and Green's function has been derived within the nonequilibrium Keldysh formalism.\cite{keldysh} This approach has been used extensively in connection with tight-binding models and first-principles methods employing localized basis sets consisting of either atomic orbitals\cite{Brandbyge,Sanvito,Taylor,sorensen} or Gaussians.\cite{gaussian} The other approach is to use a wave-function-matching method, in which the transmission and reflection coefficients of scattering wave functions are computed. This approach has been employed by combining it with techniques in which real-space grids and/or plane-wave basis sets are used to describe wave functions and potentials. \cite{lang,tsukada,choi,nkobayashi,tsukamoto,obm,obm2,icp}

When these two approaches are compared, it is advantageous to treat the charge density in the equilibrium regime of energy by the procedure of the Green's function method with the energy of a nonreal number. On the other hand, the scattering wave functions computed in a wave-function-matching method provide a direct real-space picture of the scattering process. One of the present authors has demonstrated that the wave-function-matching methods are related to the Green's function method in mathematically strict manner\cite{obm2} and can be combined straightforwardly in the real-space finite-difference (RSFD) approach,\cite{chelikowsky,tsdg,icp} which is one of the methods using real-space grids. In addition, recent cutting-edge computers worldwide rely on massively parallel architectures, and the RSFD approach is one of the most promising methods of executing large-scale simulations on massively parallel computers. Thus, it is important to develop rigorous and efficient numerical schemes combining the advantages of the Green's function and wave-function-matching methods based on the RSFD approach to perform large-scale first-principles electron-transport calculations.

In the Green's function formalism, the perturbed Green's functions of the transition region sandwiched between electrodes are computed using self-energy terms of electrodes. In our previous study,\cite{obm2} we proposed a procedure to obtain the self-energy terms from the ratio matrices, which are constituted by the generalized Bloch waves of electrodes and originally introduced in one of the wave-function-matching methods, the overbridging boundary-matching method,\cite{icp,obm} for the RSFD approach. Although the computational cost for the self-energy terms is greatly reduced, this procedure requires all the generalized Bloch waves of electrodes, and the calculation of rapidly varying evanescent waves is computationally demanding and numerically unstable.\cite{icp,obm2} S\o{}rensen {\it et al.}\cite{sorensen} proposed a method in which rapidly varying evanescent waves are excluded by introducing a cutoff for evanescent waves $\lambda_{\mbox{min}}$ in the construction of self-energy terms because rapidly varying evanescent waves included in the generalized Bloch waves do not contribute to electron transport. They demonstrated the efficiency of their method by calculating the transport properties of small molecules. However, in the case of large systems, the convergence of conductance with respect to $\lambda_{\mbox{min}}$ is slow. To perform large-scale calculations in the RSFD approach, we employ very large matrices for which self-energy terms are calculated. In addition, to avoid the problem of reduced accuracy arising from the incompleteness of the basis sets in the methods using localized basis sets, the size of the matrices is also increased. Therefore, fast and accurate computation of the self-energy terms for large systems is indispensable to ensure the reliability of transport calculations.

In this paper, we propose an efficient numerical technique to calculate the self-energy terms of electrodes for first-principles transport calculations. Using the orthogonal complement vectors of the space spanned by the Bloch waves containing the propagating waves and moderately varying evanescent waves, we can overcome the numerical difficulty in treating rapidly varying evanescent waves and in constructing self-energy terms. The computational accuracy of transport properties is significantly improved by the present technique. As an application to demonstrate the potential power of the present technique, the transport properties of a Stone-Wales (SW) defect\cite{Stone-Wales} in graphene and silicene are examined. In addition to industrial and scientific interest in two-dimensional materials with a honeycomb structure, the comparison between graphene and silicene is of importance because the electron-transport properties of an SW defect in silicene have not been intensively investigated so far in spite of its advantages over graphene. A sharp dip due to resonance scattering of the SW defect is observed in the conductance spectrum of silicene because the $\sigma^\ast$ state lies near the Fermi energy, while there are no strong dips in the spectrum of graphene in the calculated energy range. In addition, there are two conduction channels near the Fermi energy in graphene and silicene. One conduction channel is less sensitive to the SW defect at the Fermi energy while the electrons of the other channel are significantly scattered by the defect. This characteristic behavior of the transport properties of these two conduction channels is explained by the bonding arrangement of the bilattice of the honeycomb structure. Owing to the excellent accuracy of the transmission coefficients obtained using the present technique, we can distinguish the different behaviors of these two conduction channels.

The rest of this paper is organized as follows. In Sec.~\ref{sec:compmeth}, we state the problem of accuracy in transport calculations and introduce the procedure to improve accuracy together with an example demonstrating the performance of the present technique. The transport properties of an SW defect in graphene and silicene are presented in Sec.~\ref{sec:Application} and we summarize the present technique in Sec.~\ref{sec:Summary}. Finally, the transport calculation method using the RSFD approach is given in the Appendix to assist understanding of the present technique.

\section{Computational procedure to obtain self-energy terms}
\label{sec:compmeth}
\subsection{Self-energy terms from generalized Bloch waves}
\begin{figure}
\begin{center}
\includegraphics{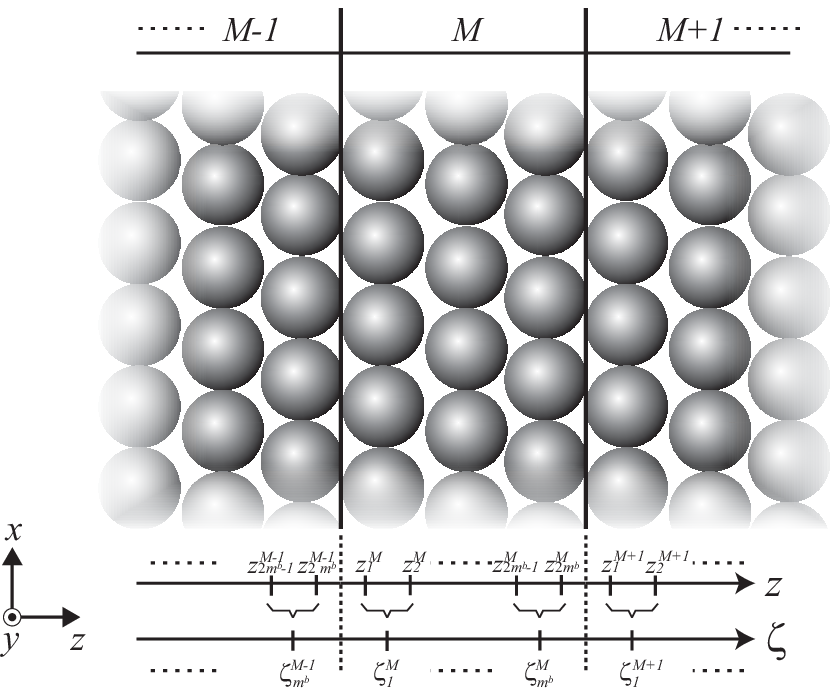}
\caption{Schematic representation of a periodic bulk. $\zeta^M_l$ represents the $z$-coordinate at the $l$th grid plane group in the $M$th unit cell. The case for $\mathcal{N}_f=2$ is illustrated as an example. \label{fig:1}}
\end{center}
\end{figure}

In this subsection, we briefly introduce the computational scheme to obtain the self-energy terms of electrodes, which is introduced in Ref.~\onlinecite{obm2}. The computational model used to obtain the generalized Bloch waves is illustrated in Fig.~\ref{fig:1}, where the atomic layers of a crystalline bulk are periodically repeated. Although we assume that electrons flow along the $z$-direction from the left to right, the opposite case can be derived in a similar manner. The system is periodic in the $x$- and $y$-directions. A generalized $z$-coordinate $\zeta_l$ is used instead of $z_l$ because a couple of grid planes are involved in the wave-function and Green's-function matching procedures when a higher-order finite-difference approximation is employed. $\zeta^M_l$ represents the $z$-coordinate at the $l$th grid plane group in the $M$th unit cell. The matching plane connecting the left (right) electrode and the transition region is between $\zeta^M_{m^b}$ ($\zeta^{M-1}_1$) and $\zeta^{M+1}_1$ ($\zeta^M_{m^b}$). In practical calculations, the order of the finite-difference approximation $\mathcal{N}_f$ is taken so as to cover the nonzero elements attributed to nonlocal parts of the pseudopotential at the matching plane and corresponds to the number of grid planes included in $\zeta_l$. The numbers of grid points in the $x$-, $y$-, and $z$-directions are $N_x$, $N_y$, and $\mathcal{N}_fm^b$, respectively. By applying the generalized Bloch condition, we can obtain the following generalized eigenvalue problem for the complex energy $Z(=E+i\eta)$:
\begin{equation}
\Pi_1(Z) \left[
\begin{array}{c}
\Phi^b_n(\zeta^{M-1}_{m^b},Z) \\
\Phi^b_n(\zeta^{M+1}_1,Z) 
\end{array}
\right] = \lambda_n(Z)\Pi_2(Z) \left[
\begin{array}{c}
\Phi^b_n(\zeta^{M-1}_{m^b},Z) \\
\Phi^b_n(\zeta^{M+1}_1,Z) 
\end{array}
\right],
\label{eqn:2-10}
\end{equation}
where
\begin{eqnarray}
\Pi_{1}(Z) 
&=&\left[
\begin{array}{cc}
 \Theta(\zeta^M_{m^b},\zeta^M_1;Z)B^b(\zeta^M_{m^b})^{\dagger}  & \: \Theta(\zeta^M_{m^b},\zeta^M_{m^b};Z)B^b(\zeta^M_{m^b}) \\
 0  &  I  \\
\end{array}
\right],  \nonumber \\
\Pi_{2}(Z) 
&=&\left[
\begin{array}{cc}
I  &  0  \\
\Theta(\zeta^M_1,\zeta^M_1;Z)B^b(\zeta^M_{m^b})^{\dagger}  & \: \Theta(\zeta^M_1,\zeta^M_{m^b};Z)B^b(\zeta^M_{m^b}) \\
\end{array}
\right].
\label{eqn:2-11}
\end{eqnarray}
Here, $\Theta(\zeta^M_k,\zeta^M_l;Z)$ is the $N(=N_x \times N_y \times \mathcal{N}_f)$-dimensional $(k,l)$ block-matrix element of the Green's function of the truncated part of the periodic Hamiltonian in the $M$th unit cell and $B^b(\zeta^M_{m^b})$ is a nonzero $N$-dimensional block-matrix element consisting of the coefficients of the finite-difference approximation and the nonlocal parts of the pseudopotential. In addition, $\Phi^b_n(\zeta^M_l,Z)$ corresponds to $N$-dimensional columnar vectors of the generalized Bloch waves at $\zeta^M_l$, $\lambda_n(Z)$ is the Bloch factor $\e^{ik_{n,z}L_z}$ with $k_{n,z}$ and $L_z$ being the $z$-component of the Bloch vector and the length of the unit cell in the $z$-direction, respectively, and the superscript $b$ indicates the matrices and vectors used to obtain the self-energy terms of the bulk electrodes. 

Now we introduce $N$-dimensional matrices $Q^{b,p}(\zeta^M_l;Z)$ and $Q^{b,q}(\zeta^M_l;Z)$, which gather the $N$ generalized Bloch waves $\{ \Phi^{b,p}_n(\zeta^M_l;Z)\}$ and $\{ \Phi^{b,q}_n(\zeta^M_l;Z)\}$, $n=1,2,...,N$, for $Z$ with $|\lambda_n| > 1$ and $|\lambda_n| < 1$, respectively:
\begin{eqnarray}
Q^{b,p}(\zeta^M_l;Z)&=&\Bigl[\Phi^{b,p}_1(\zeta^M_l;Z),\Phi^{b,p}_2(\zeta^M_l;Z),...,\Phi^{b,p}_N(\zeta^M_l;Z)\Bigr], \nonumber \\
Q^{b,q}(\zeta^M_l;Z)&=&\Bigl[\Phi^{b,q}_1(\zeta^M_l;Z),\Phi^{b,q}_2(\zeta^M_l;Z),...,\Phi^{b,q}_N(\zeta^M_l;Z)\Bigr].
\label{eqn:2-09}
\end{eqnarray}
The self-energy terms of the left and right electrodes are expressed as
\begin{eqnarray}
\Sigma^b_L(\zeta^M_l;Z)&=&B^b(\zeta^M_{l-1})^{\dagger}R^{b,p}(\zeta^M_l;Z), \nonumber \\
\Sigma^b_R(\zeta^M_l;Z) &=&B^b(\zeta^M_l) R^{b,q}(\zeta^M_{l+1};Z),
\label{eqn:2-07}
\end{eqnarray}
respectively, where $R^{b,p}(\zeta^M_l;Z)$ and $R^{b,q}(\zeta^M_l;Z)$ are the ratio matrices defined as follows:
\begin{eqnarray}
R^{b,p}(\zeta^M_l;Z)&=&Q^{b,p}(\zeta^M_{l-1};Z)Q^{b,p}(\zeta^M_l;Z)^{-1}, \nonumber \\
R^{b,q}(\zeta^M_l;Z)&=&Q^{b,q}(\zeta^M_l;Z)Q^{b,q}(\zeta^M_{l-1};Z)^{-1}.
\label{eqn:2-08}
\end{eqnarray}
According to the generalized Bloch boundary condition, it is obvious that $R^{b,A}(\zeta^M_l;Z)=R^{b,A}(\zeta^{M+1}_l;Z)$ and $Q^{b,A}_1(\zeta^{M+1}_l;Z)=\Lambda Q^{b,A}_1(\zeta^M_l;Z)$, where $A=p$ and $q$, and 
\begin{equation}
\Lambda = \left[ 
\begin{array}{ccc}
\lambda_1 & & 0 \\ 
 & \ddots &  \\ 
0 & & \lambda_N 
\end{array}
\right].
\label{eqn:2-08a}
\end{equation}

Taking the limit for the complex energy, we have
\begin{eqnarray}
\lim_{\eta \rightarrow 0^{+}} \Phi^{b,p}_n(\zeta^M_l;E + i\eta) & = & \Phi^{b,ref}_n(\zeta^M_l;E), \nonumber \\
\lim_{\eta \rightarrow 0^{+}} \Phi^{b,q}_n(\zeta^M_l;E + i\eta) & = & \Phi^{b,tra}_n(\zeta^M_l;E).
\label{eqn:2-12}
\end{eqnarray}
Then, the following matrices are obtained straightforwardly:
\begin{eqnarray}
\lim_{\eta \rightarrow 0^{+}} Q^{b,p}(\zeta^M_k,\zeta^M_l;E + i\eta) & = & Q^{b,ref}(\zeta^M_k,\zeta^M_l;E), \nonumber \\
\lim_{\eta \rightarrow 0^{+}} Q^{b,q}(\zeta^M_k,\zeta^M_l;E + i\eta) & = & Q^{b,tra}(\zeta^M_k,\zeta^M_l;E), \nonumber \\
\lim_{\eta \rightarrow 0^{+}} R^{b,p}(\zeta^M_k,\zeta^M_l;E + i\eta) & = & R^{b,ref}(\zeta^M_k,\zeta^M_l;E), \nonumber \\
\lim_{\eta \rightarrow 0^{+}} R^{b,q}(\zeta^M_k,\zeta^M_l;E + i\eta) & = & R^{b,tra}(\zeta^M_k,\zeta^M_l;E), \nonumber \\
\Sigma^{b,r}_L(\zeta^M_l;E) & = & B^b(\zeta^M_{l-1})^{\dagger} R^{b,ref}(\zeta^M_l;E), \nonumber \\
\Sigma^{b,r}_R(\zeta^M_l;E) & = & B^b(\zeta^M_l)^{\dagger} R^{b,tra}(\zeta^M_{l+1};E).
\label{eqn:2-13}
\end{eqnarray}
Here, $\Sigma^{b,r}_L(\zeta^M_l;E)$ and $\Sigma^{b,r}_R(\zeta^M_l;E)$ are the retarded self-energy term of left and right electrodes, respectively. The ratio matrix $R^{b,in}$ for the incident waves from the left electrode is introduced along similar lines into the definition of $R^{ref}$:
\begin{equation}
R^{b,in}(\zeta^M_l;E)=Q^{b,in}(\zeta^M_{l-1};E)Q^{b,in}(\zeta^M_l;E)^{-1},
\label{eqn:2-14}
\end{equation}
where
\begin{equation}
Q^{b,in}(\zeta^M_l;E)=\Bigl[\Phi^{b,in}_1(\zeta^M_{l};E),\Phi^{b,in}_2(\zeta^M_{l};E),...,\Phi^{b,in}_N(\zeta^M_{l};E) \Bigr],
\label{eqn:2-15}
\end{equation}
which includes not only ordinary right-propagating incident Bloch waves but also leftward-decreasing evanescent waves. In addition, the relationship between the retarded self-energy term $\Sigma^{b,r}_L(\zeta^M_l;E)$ and $R^{b,in}(\zeta^M_l;E)$ is
\begin{equation}
\Sigma^{b,r}_L(\zeta^M_l;E)^{\dagger} = B^b(\zeta^M_{l-1})^{\dagger} R^{b,in}(\zeta^M_l;E).
\label{eqn:2-16}
\end{equation}
Apart from the numerical difficulty in solving the generalized eigenvalue problem of Eq.~(\ref{eqn:2-10}) for rapidly varying evanescent waves, which was reported in Ref.~\onlinecite{obm}, the self-energy terms can be obtained by Eq.~(\ref{eqn:2-07}).

\subsection{Statement of the problem}
\label{sec:State of the problem}
Although the relationship between the generalized Bloch waves and self-energy terms introduced in the preceding subsection has contributed to reducing the computational cost of the self-energy terms from $O(N^3m^{b3})$ to $O(N^3)$, as shown in Ref.~\onlinecite{obm2}, the computation of the eigenvalue problems to obtain the generalized Bloch waves becomes a bottleneck when larger systems are treated. From a numerical perspective, it is convenient to compute only the generalized Bloch waves from Eq.~(\ref{eqn:2-10}) that have eigenvalues $\lambda$ within a specific interval,
\begin{equation}
|\lambda_{\mbox{min}}| \le |\lambda| \le |\lambda_{\mbox{min}}^{-1}|,
\label{eqn:3-01}
\end{equation}
for a reasonable choice of $\lambda_{\mbox{min}}$. Evanescent waves with $|\lambda_{\mbox{min}}|$ outside these regions are decaying or growing so rapidly that their contribution is negligible. The decisive factor in choosing $\lambda_{\mbox{min}}$ is that the generalized Bloch waves of electrodes must be complete in the sense that they can fully represent the transmitted and reflected waves. A couple of schemes to compute generalized Bloch waves within specific regions have been proposed: one is the Arnoldi method for the quadratic eigenvalue problem,\cite{sorensen} and the other is the Sakurai-Sugiura method\cite{sakurai-sugiura} for the generalized eigenvalue problem of Eq.~(\ref{eqn:2-10}). As long as the eigenstates are accurately calculated, the scheme used does not result in any difference in practical calculations.

When the number of generalized Bloch waves used to construct the matrix $Q^{b,p}(\zeta_l;Z)$ ($Q^{b,q}(\zeta_l;Z)$) is smaller than $N$, $Q^{b,p}(\zeta_l;Z) ^{-1}$ ($Q^{b,q}(\zeta_l;Z) ^{-1}$) in Eq.~(\ref{eqn:2-08}) is expressed as a pseudoinverse matrix and the rank of the self-energy terms is smaller than $N$. Since the rank of the coupling matrices $\Gamma_L(\zeta_0;E)$ and $\Gamma_R(\zeta_{m+1};E)$, which are the imaginary parts of the self-energy terms, is not equal to that of the perturbed Green's function $G^r_T (\zeta_k,\zeta_l;E)$ in Eq.~(\ref{eqn:2-17}), the conductance is far from that obtained using all the generalized Bloch waves in some cases. In the wave-function matching formalism, this corresponds to the problem that the number of equations in the simultaneous equations is larger than that of the unknowns, i.e., the transmission and reflection coefficients.

\begin{figure}
\begin{center}
\includegraphics{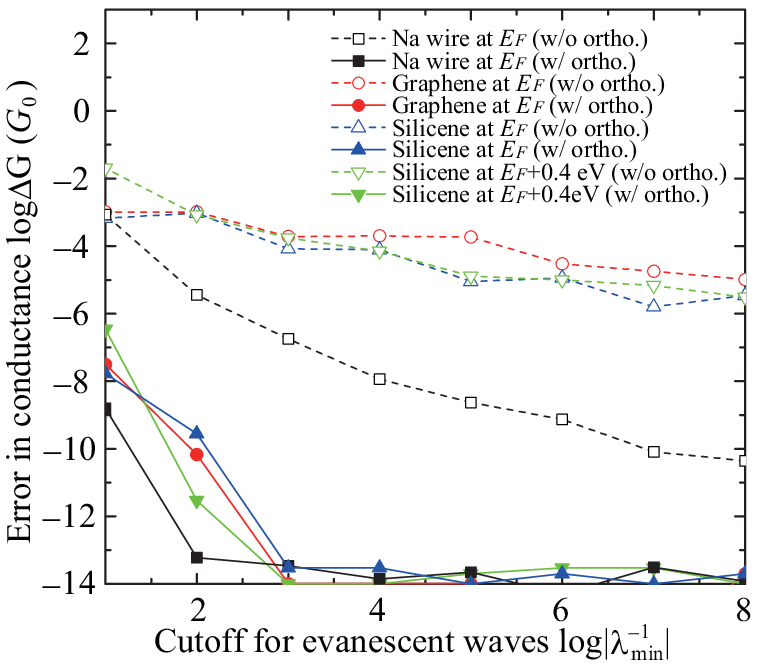}
\caption{Convergence of conductance obtained by Eq.~(\ref{eqn:2-17}) with respect to cutoff for evanescent waves. Black, red, blue, and green lines correspond to the Na wire at the Fermi energy $E_F$, graphene at $E_F$, silicene at $E_F$, and silicene at $E_F+0.4$ eV, respectively. Solid (dashed) lines indicate results obtained with (without) the orthogonal complement. \label{fig:2}}
\end{center}
\end{figure}

\begin{figure}
\begin{center}
\includegraphics{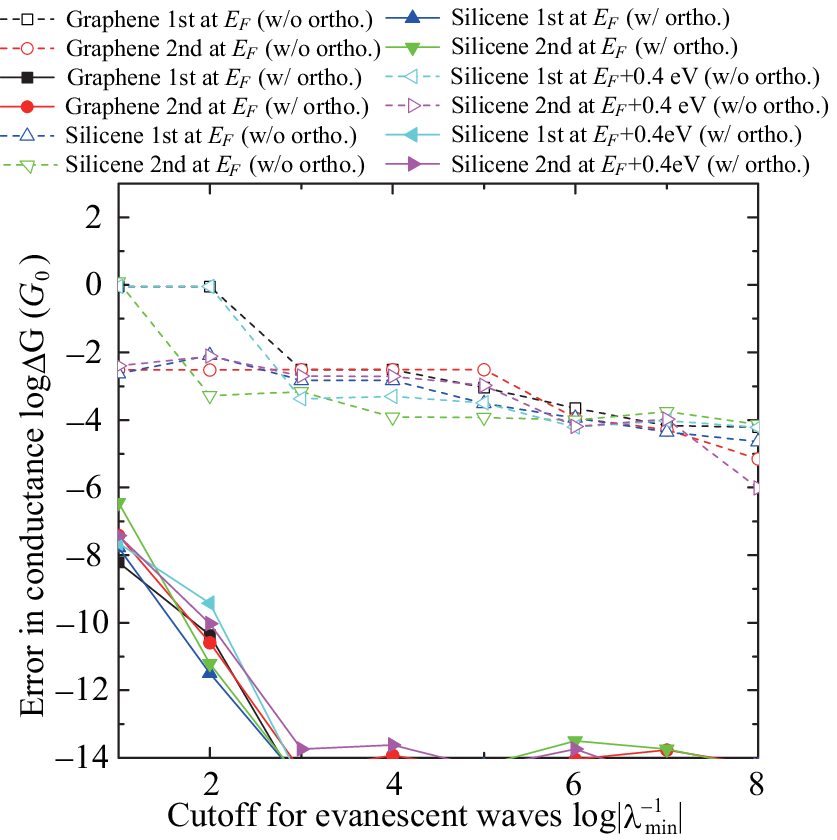}
\caption{Convergence of transmission probability of conduction channels with respect to cutoff for evanescent waves. Black, red, blue, and green lines correspond to the Na wire at the Fermi energy $E_F$, graphene at $E_F$, silicene at $E_F$, and silicene at $E_F+0.4$ eV, respectively. Solid (dashed) lines indicate results obtained with (without) the orthogonal complement. \label{fig:3}}
\end{center}
\end{figure}

To demonstrate this problem, we calculate the transport properties of a Na atomic wire, graphene, and silicene. In the case of the Na wire, the transition region contains three atoms and a grid spacing taken to be $\sim$ 0.5 bohr. The interatomic distance is 5.7 bohr and the atoms are aligned in a straight line except for the central atom of the transition region, which is replaced by an Al atom and shifted by 1.0 bohr in the direction perpendicular to the wire. The norm-conserving pseudopotentials\cite{norm} of Troullier and Martins\cite{tmpp} are employed and the exchange correlation effect is treated by the local density approximation\cite{lda} of density functional theory (DFT).\cite{dft} In the cases of graphene and silicene, an SW defect is introduced at the center of the transition region. The other computational conditions for graphene and silicene are introduced in the next section. Figures~\ref{fig:2} and \ref{fig:3} show the convergence of the conductance and transmission probability of the conduction channels with respect to $|\lambda^{-1}_{\mbox{min}}|$, respectively. Since the number of conduction channels in the Na nanowire is one, the conductance corresponds to the transmission probability of the first channel. In the case of the Na wire, the conductance converges with respect to $|\lambda^{-1}_{\mbox{min}}|$. On the other hand, in the cases of graphene and silicene, the accuracy of the conductance is not good when $|\lambda^{-1}_{\mbox{min}}|$ is not sufficiently large and the convergence of the total conductance is slow. In addition, when $|\lambda_{\mbox{min}}^{-1}|$ is large, the rapidly varying evanescent waves also cause the degradation of numerical accuracy as mentioned in the preceding subsection. Thus, the degradation of the computational accuracy due to the evanescent waves is serious when the systems become large.

The most general way to avoid this problem is to increase the thickness of buffer layers of electrodes in the transition region so that rapidly decreasing evanescent waves vanish. However, this approach is computationally demanding because the calculation of the perturbed Green's functions of the transition region is also time-consuming and requires a large amount of memory. In the next subsection, a method in which the rank of the matrices $Q^{b,p}(\zeta^M_l;Z)$ and $Q^{b,q}(\zeta^M_l;Z)$ is kept to $N$ to circumvent the problem is introduced. 

\subsection{Computational method using orthogonal complement vectors}
The generalized eigenvalue problem of Eq.~(\ref{eqn:2-10}) suffers from numerical error owing to the extremely large and small absolute values of $\lambda(Z)$ in some cases, which prevents us from accurately computing the eigenstates. To improve the accuracy of the ratio matrices, the following continued-fraction equations are introduced [see Eq.~(25) of Ref.~\onlinecite{obm}]:
\begin{widetext}
\begin{eqnarray}
R^{b,p}(\zeta^{M+1}_1;Z) &=&  \Theta(\zeta^M_{m^b},\zeta^M_{m^b};Z) B^b(\zeta^M_{m^b}) \nonumber \\
&&\hspace{-20mm} + \Theta(\zeta^M_{m^b},\zeta^M_1;Z) B^b(\zeta^M_{m^b})^\dagger \left[ R^{b,p}(\zeta^M_1;Z)^{-1} - \Theta(\zeta^M_1,\zeta^M_1;Z) B^b(\zeta^M_{m^b})^\dagger \right]^{-1} \Theta(\zeta^M_1,\zeta^M_{m^b};Z) B^b(\zeta^M_{m^b}) \nonumber \\
R^{b,q}(\zeta^{M+1}_1;Z) &=&  \Theta(\zeta^M_1,\zeta^M_1;Z) B^b(\zeta^M_{m^b})^\dagger \nonumber \\
&&\hspace{-20mm} + \Theta(\zeta^M_1,\zeta^M_{m^b};Z) B^b(\zeta^M_{m^b}) \left[ R^{b,q}(\zeta^{M+1}_1;Z)^{-1} - \Theta(\zeta^M_{m^b},\zeta^M_{m^b};Z) B^b(\zeta^M_{m^b}) \right]^{-1} \Theta(\zeta^M_{m^b},\zeta^M_1;Z) B^b(\zeta^M_{m^b})^\dagger. \nonumber \\
\label{eqn:4-01}
\end{eqnarray}
\end{widetext}
We propose a method of obtaining the self-energy terms by solving the continued-fraction equations in a self-consistent manner, the algorithm for which is given below. As shown in the Appendix, only $\Sigma^b_L(\zeta^M_1;Z)$ and $\Sigma^b_R(\zeta^M_{m^b};Z)$ are required for the transport calculation.\\
{\it Algorithm: solution of continued-fraction equations to obtain} $\Sigma^b_L(\zeta^M_1;Z)$
\begin{enumerate} 
\renewcommand{\labelenumi}{(\arabic{enumi})} 
\item Solve the generalized eigenvalue problem of Eq.~(\ref{eqn:2-10}) within the interval of $|\lambda_{\mbox{min}}|\le|\lambda|\le|\lambda_{\mbox{min}}^{-1}|$ by the Sakurai-Sugiura method.\cite{sakurai-sugiura}
\item If $\eta$ in the energy is zero, calculate the group velocity $v_g$ for the eigenstates with $|\lambda|=1$ by Eq.~(A1) of Ref.~\onlinecite{obm2}.
\item Select $\{\Phi^b_1(\zeta^M_1;Z),...,\Phi^b_K(\zeta^M_1;Z)\}$ that satisfy $\{\lambda|1<|\lambda|\le|\lambda_{\mbox{min}}^{-1}|\}\cup\{\lambda|v_g<0,|\lambda|=1\}$, where $K$ is the number of eigenstates with $\{\lambda|1<|\lambda|\le|\lambda_{\mbox{min}}^{-1}|\}\cup\{\lambda|v_g<0,|\lambda|=1\}$.
\item Prepare orthogonal complement vectors $\{\tilde{\Phi}_{K+1}(\zeta^M_1;Z),...,\tilde{\Phi}_N(\zeta^M_1;Z)\}$ of the space spanned by $\{\Phi^b_1(\zeta^M_1;Z),...,\Phi^b_K(\zeta^M_1;Z)\}$.
\item Set up $\tilde{Q}(\zeta^M_1;Z)=\Bigl[\Phi^b_1(\zeta^M_1;Z),...,\Phi^b_K(\zeta^M_1;Z),\tilde{\Phi}_{K+1}(\zeta^M_1;Z),...,\tilde{\Phi}_N(\zeta^M_1;Z)\Bigr]$.
\item Redo from (3) to (5) for $\zeta^{M-1}_{m^b}$.
\item Calculate $\tilde{R}(\zeta^M_1;Z)=\tilde{Q}(\zeta^{M-1}_{m^b};Z)\tilde{Q}(\zeta^M_1;Z)^{-1}$.
\item Solve Eq.~(\ref{eqn:4-01}) self-consistently using $\tilde{R}(\zeta^M_1;Z)$ as an initial estimate of $R^{b,p}(\zeta^M_1;Z)$.
\item Compute $\Sigma_L(\zeta^M_1;Z)$ by Eq.~(\ref{eqn:2-08}).
\end{enumerate} 
$\Sigma_R(\zeta^M_{m^b};Z)$ can be obtained in a similar manner.

Note that $\tilde{R}(\zeta^M_1;Z)$ is a good initial estimate for $R^{b,p}(\zeta^M_1;Z)$. Since the rank of $\tilde{Q}(\zeta^M_l;Z)$ is equal to that of $Q^{b,p}(\zeta^M_l;Z)$, the eigenstates $\{\Phi_{K+1}(\zeta^M_l;Z),...,\Phi_N(\zeta^M_l;Z)\}$ outside the intervals $|\lambda_{\mbox{min}}| \le |\lambda| \le |\lambda_{\mbox{min}}^{-1}|$ can be described by a linear combination of the $N$-dimensional columnar vectors consisting of $\tilde{Q}(\zeta^M_l;Z)$.
\begin{eqnarray}
Q^{b,p}(\zeta^{M-1}_{m^b};Z)&=&\tilde{Q}(\zeta^{M-1}_{m^b};Z)P(\zeta^{M-1}_{m^b};Z), \nonumber \\
Q^{b,p}(\zeta^M_1;Z)&=&\tilde{Q}(\zeta^M_1;Z)P(\zeta^M_1;Z),
\label{eqn:4-02}
\end{eqnarray}
where $P(\zeta^{M-1}_{m^b};Z)$ and $P(\zeta^M_1;Z)$ are matrices composed of the coefficients of the linear combination. When the grid spacing in the $z$-direction is sufficiently small, $P(\zeta^{M-1}_{m^b};Z) \approx P(\zeta^M_1;Z)$. Thus, we have
\begin{eqnarray}
\tilde{R}(\zeta^M_1;Z)&=&\tilde{Q}(\zeta^{M-1}_{m^b};Z) \tilde{Q}(\zeta^M_1;Z)^{-1} \nonumber \\
&=&Q(\zeta^{M-1}_{m^b};Z) P(\zeta^{M-1}_{m^b};Z) P(\zeta^M_1;Z)^{-1}Q^{b,p}(\zeta^M_1;Z)^{-1} \nonumber \\
&\approx&R^{b,p}(\zeta^M_1;Z).
\label{eqn:4-03}
\end{eqnarray}

\subsection{Numerical test}
To examine the efficiency of the present technique, we examined the transport properties of the systems considered in Sec.~\ref{sec:State of the problem}. Figures~\ref{fig:2} and \ref{fig:3} also show the convergence of the conductance and transmission probability of the conduction channels obtained by the present technique as solid lines. The convergence is much faster than that obtained without the orthogonal complement vectors. According to the results obtained using the orthogonal complement vectors, evanescent waves within the interval of $10^{-3}\le |\lambda| \le 10^3$ affect the transport properties when the double precision of Fortran 95 is employed. The numerical error in the scheme without the orthogonal complement vectors is caused by the use of pseudoinverse matrices. In addition, by solving the continued-fraction equations, Eq.~(\ref {eqn:4-01}), the degradation of the numerical accuracy is suppressed when $|\lambda_{\mbox{min}}^{-1}|$ is large. Thus, we can conclude that the present technique significantly improves the convergence with respect to the cutoff of evanescent waves for large systems.

\section{Application}
\label{sec:Application}
Graphene,\cite{graphene} in which $sp^2$ hybridized electrons ($\sigma$ electrons) form a honeycomb structure and the remaining $\pi$ ($p_z$) electrons follow the massless Dirac equations, has attracted a great deal of interest. Owing to its unique structural and electronic properties, graphene is as an important material for numerous theoretical investigations and a promising material for applications. Although the research interest in graphene is growing rapidly, there is increasing interest in whether the other group IV elements in the periodic table have a stable honeycomb structure. DFT has revealed that silicene, which is a honeycomb structure of Si, is stable in the form of a slightly buckled structure in which the neighboring atoms are alternately displaced perpendicular to the plane and $p_z$ electrons behave as massless Dirac fermions.\cite{takeda,APL97_163114} Recently, the possible growth of silicene on a Ag(110) or Ag(100) substrate has been reported.\cite{silicene1,silicene2,silicene3} Although silicene has advantages over graphene because of its high compatibility with current Si-based device technologies, few theoretical studies on the transport properties of defects in silicene have been conducted so far.

Regarding the defects in graphene, the controllable defects mainly include SW defects,\cite{Stone-Wales} adatoms, vacancies, substitution, and disorder. Among them, SW defects are important topological defects in materials with a honeycomb structure, playing a central role in their formation, transformation, fracture, and embrittlement. SW defects are also expected to alter the electronic structures of graphene and affect its unique transport properties.\cite{grapheneSW1,grapheneSW2} Owing to the weak overlapping between the $p_z$ orbitals between neighbor atoms in silicene, gaining a basic knowledge of SW defects in silicene is essential to deepen fundamental understanding of the transport properties of materials with a honeycomb structure.

\begin{figure}
\begin{center}
\includegraphics{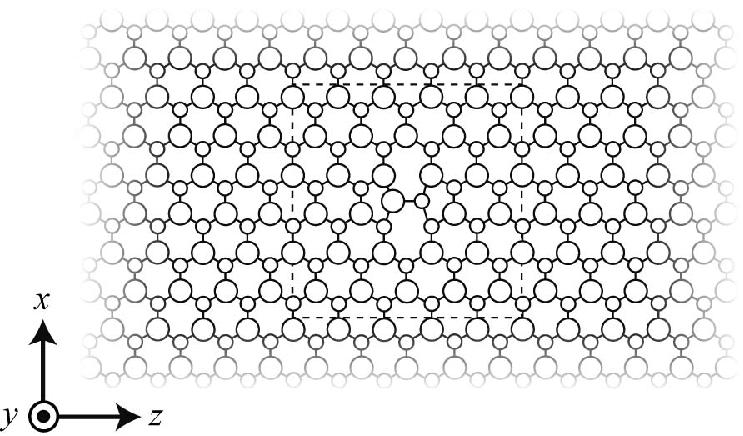}
\caption{Schematic image of computational model. Large and small circles represent the upper and lower atoms of an alternately buckled honeycomb structure. In the case of graphene, the lattice is not alternately buckled, but slightly wavy. \label{fig:4}}
\end{center}
\end{figure} 

\begin{figure}
\begin{center}
\includegraphics{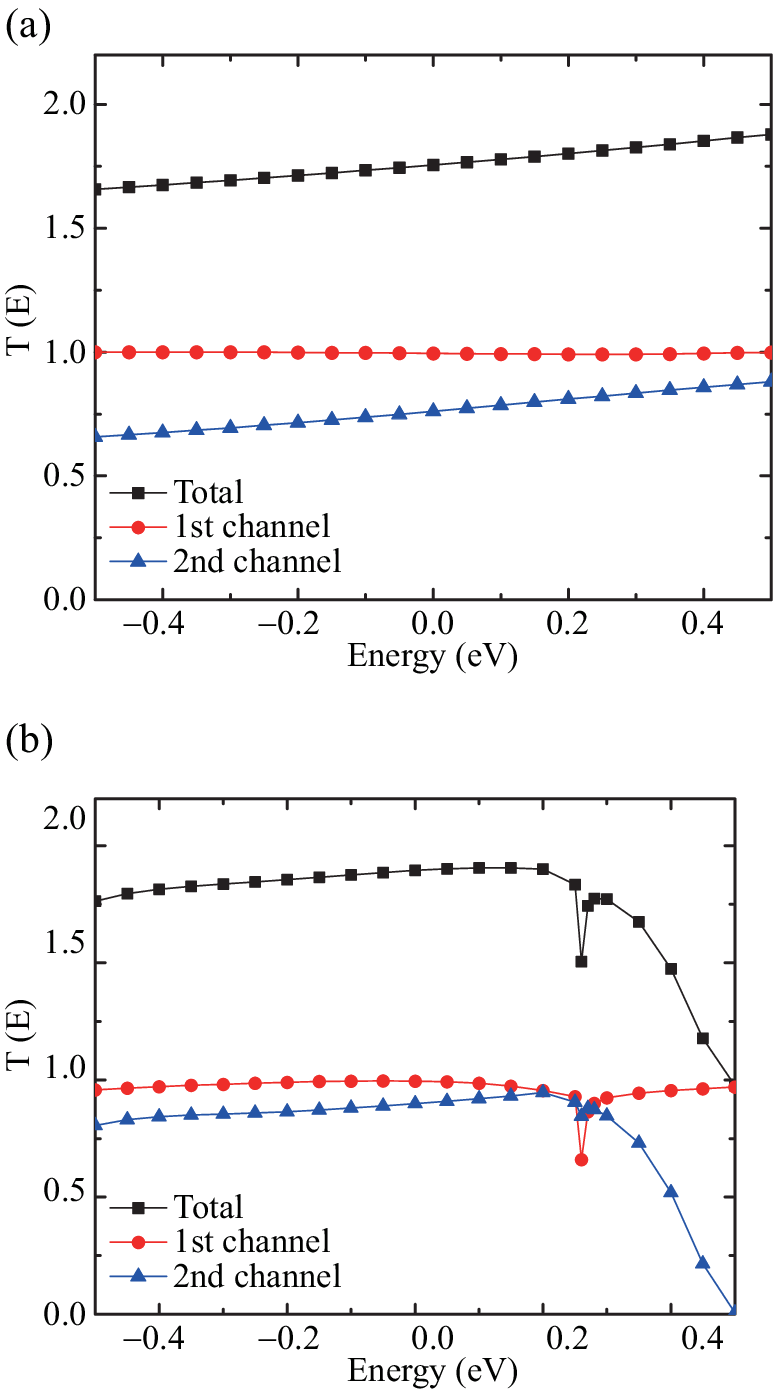}
\caption{Conductance and transmission probabilities of conduction channels for (a) graphene and (b) silicene. \label{fig:5}}
\end{center}
\end{figure} 

\begin{figure}
\begin{center}
\includegraphics{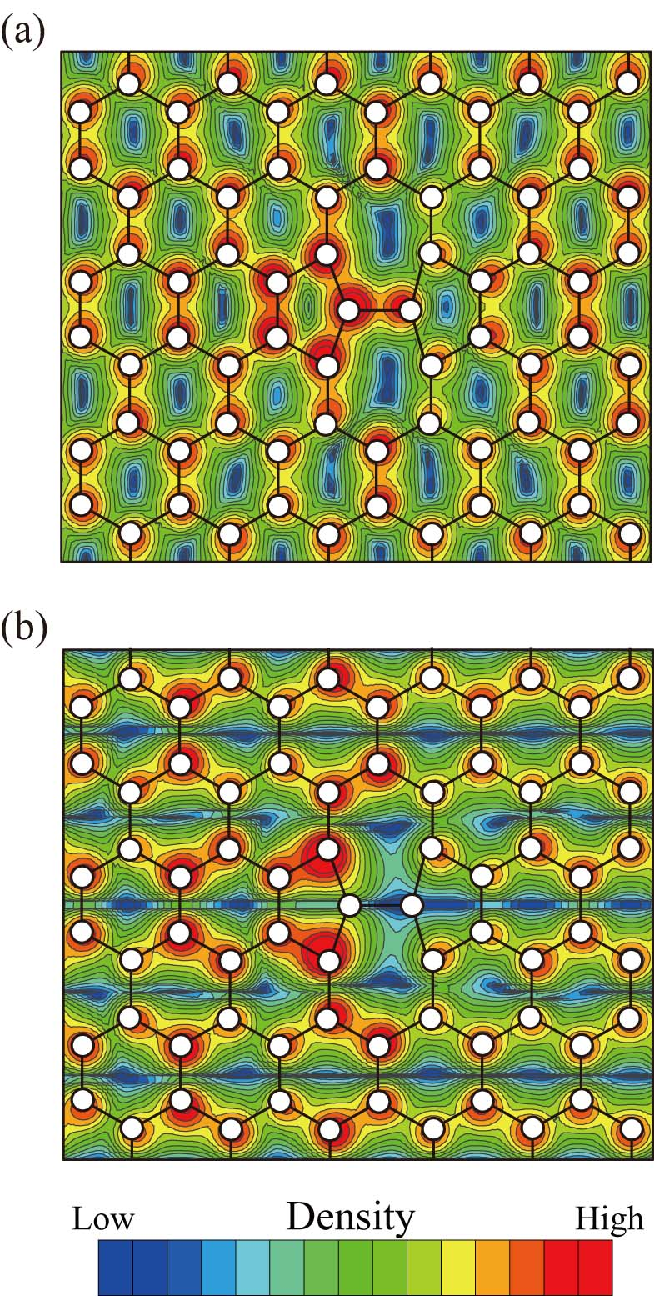}
\caption{Charge density distribution of scattering wave functions of graphene. (a) First channel and (b) second channel. Each contour represents twice or half
the density of the adjacent contour lines, and the lowest contour is $2.48 \times 10^{-7}$ electron/eV/\AA$^3$. \label{fig:6}}
\end{center}
\end{figure}

\begin{figure}
\begin{center}
\includegraphics{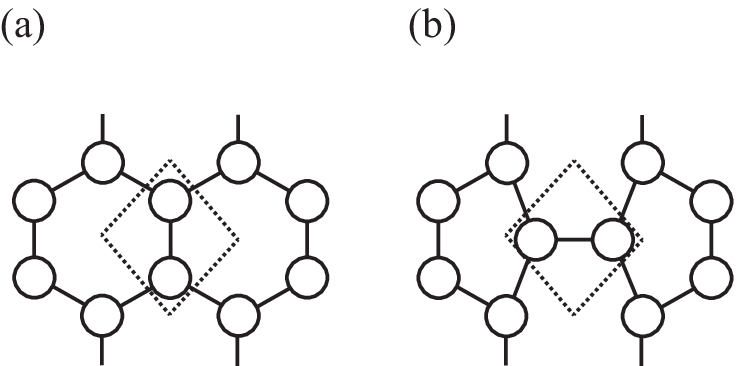}
\caption{Schematic image of formation of SW defect. (a) Before deformation and (b) after deformation. The dotted line indicates a unit cell of the bilattice. \label{fig:7}}
\end{center}
\end{figure}

First, we examine the optimized atomic structures of graphene and silicene without any defects as well as with an SW defect. The grid spacing is set at $\sim$ 0.33 and $\sim$ 0.40 bohr for graphene and silicene, respectively. Integration over the Brillouin zone is carried out using equidistant $k$-point sampling, in which the $k$-point density is chosen so as to correspond to 144-point sampling in the irreducible Brillouin zone of pristine graphene and silicene. The exchange-correlation effect is treated by the local density approximation\cite{lda} of DFT, and the projector augmented wave method\cite{paw} is used to describe the electron-ion interaction. The supercell obtaining the atomic structure of the SW defect is indicated by the dashed line in Fig.~\ref{fig:4}. In the case of graphene, the lattice is not alternately buckled, but slightly wavy.\cite{Ma} The cutoff for the evanescent waves $|\lambda_{\mbox{min}}|$ is set to $10^{-3}$. We then examine the transport properties by embedding the transition region with an SW defect enclosed by the dashed line in Fig.~\ref{fig:4} in the honeycomb structure. In the transport calculation performed to obtain the scattering wave functions, the norm-conserving pseudopotentials\cite{norm} of Troullier and Martins\cite{tmpp} are employed instead of the projector augmented wave method. To determine the Kohn-Sham effective potential, a supercell is used under a periodic boundary condition, and then the scattering wave functions are computed under the semi-infinite boundary condition obtained non-self-consistently. It has been reported that this procedure is just as accurate in the linear response regime but significantly more efficient than performing computations self-consistently on a scattering-wave basis.\cite{kong2} Figure~\ref{fig:5} shows the total conductance and transmission probability of the conduction channels. Two conduction channels contribute to electron transport at $k_x=0$ while there are no conduction channels at other points. A strong dip is observed at $E_F+0.26$ eV in the case of silicene. By plotting the charge density distribution of the scattering wave (not shown), it was found that the dip can be ascribed to resonance scattering of the SW defect. The $\sigma^\ast$ state of pristine silicene is 1.2 eV above the Fermi energy while that of graphene is 3.5 eV above the Fermi energy.\cite{Graphene_2_74} The SW defect of silicene scatters electrons with energy slightly above the Fermi energy. Furthermore, the transmission probability of the first channel is almost unity near the Fermi energy while that of the second is considerably reduced by the scattering at the SW defect. The charge density distribution of the scattering wave functions in graphene are plotted in Fig.~\ref{fig:6}. The charge density of the first channel accumulates so as to connect carbon atoms in the $x$-direction while that of the second channel aligns along the $z$-direction. The first channel forms a bond between carbon atoms in the same unit cell of the bilattice, while the second channel connects carbon atoms in the neighboring unit cells. The SW defect is formed by rotating a carbon-carbon bond in the unit cell as shown in Fig.~\ref{fig:7}. The second channel is easily affected because the bond configuration is greatly deformed. This feature is also observed in silicene. Owing to the accurate evaluation of the transmission probability obtained by the present technique, the transport properties of the first and second channels can be distinguished.

\section{Summary}
\label{sec:Summary}
We have developed a numerical technique to obtain the self-energy terms of electrodes for first-principles transport calculations. The present method can significantly improve the computational accuracy of transport properties without using all the generalized Bloch waves of electrodes. The self-energy terms of electrodes are computed using the generalized Bloch waves that actually contribute to transport phenomena and the orthogonal complement vectors of the space spanned by the Bloch waves. By solving the continued-fraction equations developed in the overbridging boundary-matching method,\cite{obm,obm2,icp} we obtain the self-energy terms with high degrees of accuracy. In addition, the present technique is particularly efficient for large-scale transport simulations employing the RSFD approach because the matrix size is taken to be large so as to perform highly accurate calculations.

To present the efficiency of the present technique, the electron-transport properties of an SW defect in graphene and silicene are calculated. Since the $\sigma^\ast$ state of pristine silicene lies at a lower energy than that of graphene, a sharp dip ascribed to resonance scattering of the SW defect is observed in the conductance spectrum of silicene. In addition, there are two conduction channels near the Fermi energy in graphene and silicene. One conduction channel is easily affected by the SW defect while the other is insensitive at the Fermi energy. The deformation of the bonding network, which connects the bilattices of the honeycomb structure in the formation of SW defects, causes this characteristic difference in the transport properties of these two channels. Owing to the excellent computational accuracy of the present technique, the different behaviors of these two conduction channels in graphene and silicene can be distinguished.

The RSFD scheme for first-principles calculations has the advantage of potential scaling with massively parallel architectures without compromising on accuracy. From the above, it seems reasonable to conclude that the present technique opens the possibility of executing large-scale transport calculations using massively parallel computers.

\section*{Acknowledgements}
This research was partially supported by the Computational Materials Science Initiative (CMSI), and by a Grant-in-Aid for Scientific Research on Innovative Areas (Grant No. 22104007) from the Ministry of Education, Culture, Sports, Science and Technology, Japan. The numerical calculation was carried out using the computer facilities of the Institute for Solid State Physics at the University of Tokyo and Center for Computational Sciences at University of Tsukuba.

\appendix
\section{Transport calculation method using density functional theory}
\label{sec:method}
\subsection{Perturbed Green's function and self-energy term}
\begin{figure*}
\begin{center}
\includegraphics{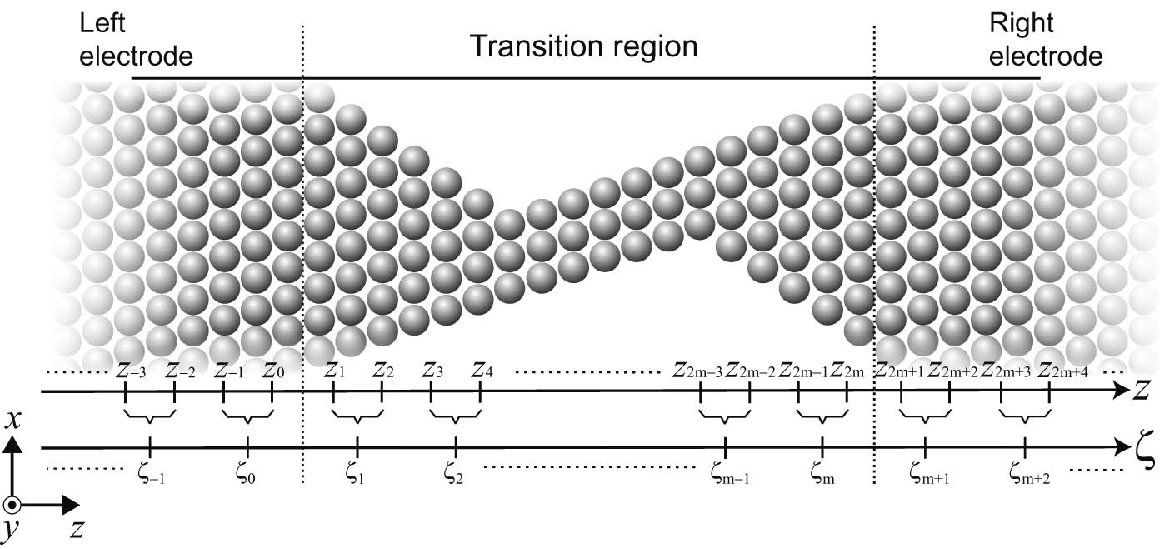}
\caption{Sketch of a system with a transition region intervening between left and right semi-infinite crystalline electrodes. The dotted lines correspond to the borders of the partitioning of the Hamiltonian matrix in Eq.~(\ref{eqn:2-01}) and Fig.~\ref{fig:9}. The case for $\mathcal{N}_f=2$ is illustrated as an example. \label{fig:8}}
\end{center}
\end{figure*}

In this Appendix, we briefly summarize the procedure used to compute the perturbed Green's functions and transport properties using the self-energy terms of electrodes for convenience. Since the proof has already been reported in Refs.~\onlinecite{obm} and \onlinecite{obm2}, here we introduce important formulae to explain the newly developed technique. A typical computational model for a transport calculation is illustrated in Fig.~\ref{fig:8}, where a nanostructure is sandwiched between semi-infinite electrodes. Two-dimensional periodicity in the $x$- and $y$-directions is assumed and a generalized $z$-coordinate $\zeta_l$ is used instead of $z_l$. The exchange-correlation effect is treated by the local density approximation\cite{lda} or generalized gradient approximation\cite{gga} of DFT\cite{dft} to neglect the interaction between the left and right electrodes.
\begin{figure}
\begin{center}
\includegraphics{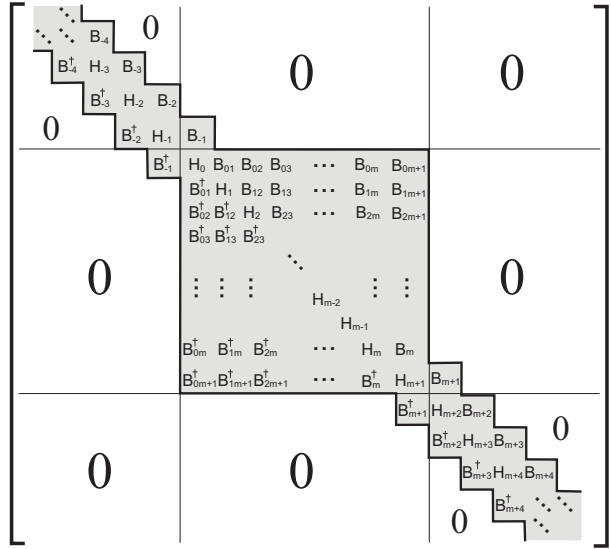}
\caption{Partitioning of the Hamiltonian matrix $\hat{H}$ of Eq.~(\ref{eqn:2-01}). Block-matrix elements $H_l$, $B_l$, and $B_{ll'}$ are abbreviations of $H(\zeta_l,\vecvar{k}_{||})$, $B(\zeta_l)$, and $B(\zeta_l,\zeta_{l'})$, respectively. The partition lines are identical to those in Eq.~(\ref{eqn:2-01}). Reprinted from Ref.~\onlinecite{obm2}. \label{fig:9}}
\end{center}
\end{figure}

As shown in Fig.~\ref{fig:9}, we are interested in the finite part of the Kohn-Sham Hamiltonian\cite{kohnsham} matrix,
\begin{equation}
\hat{H}(\vecvar{k}_{||})=
\left[ 
\begin{array}{c|c|c}
\hat{H}_L(\vecvar{k}_{||}) & \hat{B}_{LT} & 0 \\ 
\hline
 \hat{B}_{LT}^{\dagger} & \hat{H}_T(\vecvar{k}_{||}) & \hat{B}_{TR} \\ 
\hline
0 & \hat{B}_{TR}^{\dagger} & \hat{H}_R(\vecvar{k}_{||})
\end{array}
\right],
\label{eqn:2-01}
\end{equation}
where the borders of the partitioning of $\hat{H}(\vecvar{k}_{||})$ are drawn as dotted lines in Fig.~\ref{fig:8}; the submatrix $\hat{H}_T(\vecvar{k}_{||})$ contains the matrix elements in the transition region, $\hat{H}_L(\vecvar{k}_{||})$ ($\hat{H}_R(\vecvar{k}_{||})$) corresponds to the semi-infinite left (right)-electrode region, and $\hat{B}_{LT}$ ($\hat{B}_{TR}$) is the coupling term between the transition region and the left (right) electrode. The perturbed Green's function, which contains the effect of the electrodes, is defined as
\begin{eqnarray}
\hat{G}(Z,\vecvar{k}_{||})&=& \left[ Z-\hat{H}(\vecvar{k}_{||}) \right]^{-1} \nonumber \\
&=& \left[ 
\begin{array}{c|c|c}
\hat{G}_L(Z,\vecvar{k}_{||}) & \hat{G}_{LT}(Z,\vecvar{k}_{||}) & \hat{G}_{LR}(Z,\vecvar{k}_{||}) \\ \hline
\hat{G}_{TL}(Z,\vecvar{k}_{||}) & \hat{G}_T(Z,\vecvar{k}_{||}) & \hat{G}_{TR}(Z,\vecvar{k}_{||}) \\ \hline
\hat{G}_{RL}(Z,\vecvar{k}_{||}) & \hat{G}_{RT}(Z,\vecvar{k}_{||}) & \hat{G}_R(Z,\vecvar{k}_{||})
\end{array}
\right].
\label{eqn:2-02}
\end{eqnarray}
From the matrix equation
\begin{equation}
\left[ 
\begin{array}{ccc}
Z-\hat{H}_L(\vecvar{k}_{||}) & -\hat{B}_{LT} & 0 \\ 
-\hat{B}_{LT}^{\dagger} & Z-\hat{H}_T(\vecvar{k}_{||}) & -\hat{B}_{TR} \\ 
0 & -\hat{B}_{TR}^{\dagger} & Z-\hat{H}_R(\vecvar{k}_{||})
\end{array}
\right]
\left[ 
\begin{array}{l}
\hat{G}_{LT}(Z,\vecvar{k}_{||}) \\ 
\hat{G}_{T}(Z,\vecvar{k}_{||}) \\ 
\hat{G}_{RT}(Z,\vecvar{k}_{||}) 
\end{array}
\right]
=
\left[ 
\begin{array}{c}
0 \\ 
I \\ 
0 
\end{array}
\right],
\label{eqn:2-03}
\end{equation}
we obtain
\begin{eqnarray}
\hat{G}_{LT}(Z,\vecvar{k}_{||})\hat{G}_T(Z,\vecvar{k}_{||})^{-1} = \left[ Z-\hat{H}_L(\vecvar{k}_{||}) \right]^{-1} \hat{B}_{LT}, \nonumber \\
-\hat{B}_{LT}^{\dagger}\hat{G}_{LT}(Z,\vecvar{k}_{||}) + \left[ Z-\hat{H}_{T}(\vecvar{k}_{||}) \right] \hat{G}_T(Z,\vecvar{k}_{||})-\hat{B}_{TR}\hat{G}_{RT}(Z,\vecvar{k}_{||})=I, \nonumber \\
\hat{G}_{RT}(Z,\vecvar{k}_{||})\hat{G}_T(Z,\vecvar{k}_{||})^{-1} = \left[ Z-\hat{H}_R(\vecvar{k}_{||}) \right]^{-1}\hat{B}_{TR}^{\dagger}. \nonumber \\
\label{eqn:2-04}
\end{eqnarray}
One sees that the perturbed Green's function $\hat{G}_T(Z,\vecvar{k}_{||})$ can be portioned to the transition region as
\begin{equation}
\hat{G}_T(Z,\vecvar{k}_{||}) =\left[ Z-\hat{H}_T(\vecvar{k}_{||})- \hat{\Sigma}_L(Z,\vecvar{k}_{||})-\hat{\Sigma}_R(Z,\vecvar{k}_{||}) \right]^{-1}
\label{eqn:2-05}
\end{equation}
with $\hat{\Sigma}_L(Z,\vecvar{k}_{||})$ and $\hat{\Sigma}_R(Z,\vecvar{k}_{||})$ being the self-energy terms of the left and right electrodes, respectively. Note that Eq.~(\ref{eqn:2-05}) is equivalent to Dyson's equation in the standard form.\cite{dyson} In the RSFD scheme, $\hat{B}_{LT}$ ($\hat{B}_{TR}$) has only one nonzero $N(=N_x \times N_y \times \mathcal{N}_f)$-dimensional block-matrix element $B(\zeta_{-1})$ ($B(\zeta_{m+1})$), which corresponds to $B^b(\zeta^M_{m^b})$ for the left electrode ($B^b(\zeta^M_1)$ for the right electrode), as illustrated in Fig.~\ref{fig:9}. The self-energy terms are found to take the very simple form of 
\begin{eqnarray}
\hat{\Sigma}_L(Z) & = &
\left[ 
\begin{array}{cccc}
\Sigma_L(\zeta_0;Z) & 0 & \cdots & 0 \\ 
0 & 0 & \cdots & 0 \\ 
\vdots &  & \vdots &  \\ 
0 & 0 & \cdots & 0 
\end{array}
\right] \nonumber\\
\hat{\Sigma}_R(Z) & = &
\left[ 
\begin{array}{cccc}
0 & \cdots & 0 & 0 \\ 
  & \vdots &   & \vdots   \\ 
0 & \cdots & 0 & 0 \\ 
0 & \cdots & 0 & \Sigma_R(\zeta_{m+1};Z) 
\end{array}
\right],
\label{eqn:2-06}
\end{eqnarray}
where $\Sigma_L(\zeta_{0};Z)=\Sigma^b_L(\zeta^M_{m^b};Z)$ and $\Sigma_R(\zeta_{m+1};Z)=\Sigma^b_R(\zeta^M_{1};Z)$ in Sec. II.

Conductance is calculated by the following well-known formula\cite{fisherlee} in the NEGF formalism pioneered by Keldysh:\cite{keldysh}
\begin{widetext}
\begin{equation}
G(E) = \frac{2e^2}{h} \mbox{Tr} \Bigl[ \Gamma_L(\zeta_0;E)G^r_T(\zeta_0,\zeta_{m+1};E)^{\dagger}\Gamma_R(\zeta_{m+1};E)G^r_T(\zeta_{m+1},\zeta_0;E) \Bigr],
\label{eqn:2-17}
\end{equation}
\end{widetext}
where
\begin{equation}
G^r_T (\zeta_k,\zeta_l;E) = \lim_{\eta \rightarrow 0^{+}} G_{T} (\zeta_k,\zeta_l;E + i\eta)
\label{eqn:2-18}
\end{equation}
and
$\Gamma_L$ ($\Gamma_R$) is the coupling matrix, which describes the `coupling strength' of the transition region to the left (right) electrode at $\zeta_0$ ($\zeta_{m+1}$), and is defined by
\begin{eqnarray}
\Gamma_L(\zeta_0;E)&=&i\Bigl[ \Sigma_L(\zeta_0;E)- \Sigma_L(\zeta_0;E)^{\dagger} \Bigr],\nonumber \\
\Gamma_R(\zeta_{m+1};E)&=&i\Bigl[ \Sigma_R(\zeta_{m+1};E)- \Sigma_R(\zeta_{m+1};E)^{\dagger} \Bigr].
\label{eqn:2-19}
\end{eqnarray}

Scattering wave functions in the electrodes are expressed as
\begin{equation}
\Psi_j(\zeta_l;E)=
\left\{
\begin{array}{lcl}
\displaystyle{
\Phi_j^{in}(\zeta_l;E)+\sum_{n=1}^Nr_{ij}\Phi_n^{ref}(\zeta_l;E)} & \cdots & l\leq0 \\
\displaystyle{
\sum_{n=1}^Nt_{ij}\Phi_n^{tra}(\zeta_l;E)} & \cdots & l\geq m+1
\end{array}
\right.
\label{eqn:2-20}
\end{equation}
with $t_{ij}$ and $r_{ij}$ being transmission and reflection coefficients, respectively. The transmission and reflection coefficients are computed using
\begin{eqnarray}
T &=& i Q^{tra}(\zeta_{m+1};E)^{-1}G^r_T (\zeta_{m+1},\zeta_0;E) \Gamma_L(\zeta_0;E)Q^{in}(\zeta_0;E), \nonumber \\
R &=& i Q^{ref}(\zeta_0;E)^{-1} G^r_T (\zeta_0,\zeta_0;E) \Gamma_L(\zeta_0;E) - Q^{ref}(\zeta_0;E)^{-1} Q^{in}(\zeta_0;E), \nonumber \\
\label{eqn:2-21}
\end{eqnarray}
where $T$ and $R$ are the transmission-coefficient and reflection-coefficient matrices, which are given as
\begin{equation}
T=\left[
\begin{array}{cccc}
t_{11} & t_{12} & \cdots & t_{1N} \\
t_{21} & t_{22} & \cdots & t_{2N} \\
       & \cdots &        &        \\
t_{N1} & t_{N2} & \cdots & t_{NN} \\
\end{array}
\right]
\label{eqn:2-22}
\end{equation}
and
\begin{equation}
R=\left[
\begin{array}{cccc}
r_{11} & r_{12} & \cdots & r_{1N} \\
r_{21} & r_{22} & \cdots & r_{2N} \\
       & \cdots &        &        \\
r_{N1} & r_{N2} & \cdots & r_{NN} \\
\end{array}
\right] ,
\label{eqn:2-23}
\end{equation}
respectively. Here, $\Phi_n^{in}(\zeta_l;E)$ ($Q^{in}(\zeta_l;E)$), $\Phi_n^{ref}(\zeta_l;E)$ ($Q^{ref}(\zeta_l;E)$), and $\Phi_n^{tra}(\zeta_l;E)$ ($Q^{tra}(\zeta_l;E)$) are $\Phi_n^{b,in}(\zeta^M_l;E)$ ($Q^{b,in}(\zeta^M_l;E)$) for the left electrode, $\Phi_n^{b,ref}(\zeta^M_l;E)$ ($Q^{b,ref}(\zeta^M_l;E)$) for the left electrode, and $\Phi_n^{b,tra}(\zeta^M_{l-m};E)$ ($Q^{b,tra}(\zeta^M_{l-m};E)$) for the right electrode, respectively. Note that $Q^{b,tra}$ and $Q^{b,ref}$ include the orthogonal complement vectors in the technique introduced in Sec.~\ref{sec:compmeth}. Conductance is also calculated as
\begin{equation}
G(E) = \frac{2e^2}{h} \sum_{i,j} |t_{ij}|^2 \frac{v'_i}{v_j},
\label{eqn:2-24}
\end{equation}
where $v'_i$ and $v_j$ are the group velocities of incident and transmitted propagating waves, which are defined in Ref.~\onlinecite{obm2}.

\end{document}